\documentclass[10pt,aps,pre,twocolumn,notitlepage,superscriptaddress,preprintnumbers]{revtex4-1}
\usepackage[utf8]{inputenc}
\usepackage[utf8]{inputenc} 
\usepackage[T1]{fontenc}
\usepackage{amsmath,amssymb,amsfonts}
\usepackage{bm}
\usepackage{graphicx,}
\usepackage{verbatim}
\usepackage{float}
\usepackage[FIGTOPCAP]{subfigure} 
\usepackage{dcolumn}
\usepackage{natbib}
\usepackage{hyperref}
\usepackage{tikz}
\usetikzlibrary {arrows.meta} 
\usetikzlibrary{patterns}
\usetikzlibrary{hobby}
\usepackage{appendix}
\usepackage{ulem}
\usepackage[dvipsnames]{xcolor}
\usepackage{amssymb}

\newcommand{\RN}[1]{%
	\textup{\uppercase\expandafter{\romannumeral#1}}%
}
\begin{document}
	\title{A Statistical Analysis of Diffusion Dynamics in Networks}
 	\author{A. Talebi}
	\email{ali.talebi.physics@gmail.com}
	\affiliation{Department of Physics, Shahid Beheshti University, Evin, Tehran 1983969411, Iran}

	\date{\today}
\begin{abstract}
Network diffusion is a well-known model for studying how a quantity in a network propagates to reach a steady state, where its value becomes uniformly distributed across all nodes. Previous studies of network diffusion have mainly focused on the second-smallest eigenvalue of the network Laplacian to determine the timescale associated with the slowest decaying mode. However, this study investigates the eigenspectrum of the network Laplacian in terms of its statistical properties. This method is useful for studying the statistical properties of network diffusion without calculating all eigenvalues, which can be useful for large networks. Additionally, this study introduces a condition under which the system is less diffusive and investigates how changes in the network structure influence the overall diffusion dynamics. 
\end{abstract}
\maketitle
\section{Introduction}
The study of networks has grown to model complex systems \cite{albert2002statistical,barabasi1999emergence,barabasi2013network,boccaletti2006complex,newman2018networks}. Many real‑world systems can be drawn as networks, where the nodes are interacting entities and the edges show how they interact. A network can represent several natural \cite{jeong2000large,jeong2001lethality}, social \cite{milgram1967small,watts1998collective}, and technological systems \cite{faloutsos1999power,jeong1999diameter}. One of the most important goals of studying networks is a deeper understanding of network structure. However, the structure and properties of a network are not always static, and they can evolve. Studying diffusion on networks provides a framework to study the dynamics of a quantity in a network \cite{mohar1991laplacian}. The quantities that evolve in a network can be signals in neural systems \cite{raj2012network, abdelnour2014network}, diseases in a community \cite{pastor2015epidemic, keeling2005networks}, and information in a technological system \cite{gomez2012inferring}. Studying the effect of a network's structure on diffusion dynamics helps us predict transport efficiency, disease spread, and information flow in a complex network. 

A wide range of linear and nonlinear models \cite{watts2002simple, nakao2010turing, castellano2009nonlinear} have been developed to describe diffusion in a complex network. Researchers have studied diffusion models in networks with different structures to learn how network structure, such as sparsity \cite{gomez2008entropy}, degree distribution \cite{pastor2001epidemic}, clustering \cite{serrano2006}, modularity \cite{nematzadeh2014optimal}, and multilayer organization \cite{gomez2013diffusion}, influences diffusion speed. Beyond undirected networks, recent studies focus on diffusion in complicated models, such as a model where edge directionality and multilayer organization add complexity to the network \cite{bouchet2026directionality}. The models with additional complexity were helpful for better analyzing real-world systems such as the London transport system \cite{bouchet2026directionality}. 

The simplest and well-known model of diffusion in networks is the linear diffusion model
\begin{equation}\label{EQ:1}
\frac{d x_i}{d t} = D \sum_{j=1}^{N} w_{ij}\left(x_j-x_i\right),
\end{equation}
where the $w_{ij}$ is the weight matrix, and $D$ is the diffusion constant. According to Eq.(\ref{EQ:1}), the rate of change of the state of each node is proportional to the difference between its state and those of its neighboring nodes. Additionally, this rate depends on the weight matrix and the diffusion constant. The equation Eq.(\ref{EQ:1}) can be simplified as
\begin{equation}\label{EQ:2}
\frac{d\mathbf{x}}{dt} = -D\mathcal{L}\mathbf{x},
\end{equation}
where $\mathbf{x}$ is the vector contains all nodes, and $\mathcal{L} \in \mathbb{R}^{N \times N}$ is known as Laplacian matrix is defined as
\begin{equation}\label{EQ:3}
\mathcal{L}=
\left(
\begin{array}{ccccc}
a_{1} & -w_{12} & -w_{13} & \cdots & -w_{1N}\\
-w_{21} & a_{2} & -w_{23} & \cdots & -w_{2N}\\
-w_{31} & -w_{32} & a_{3} & \cdots & -w_{3N}\\
\vdots & \vdots & \vdots & \ddots & \vdots\\
-w_{N1} & -w_{N2} & -w_{N3} & \cdots & a_{N}
\end{array}
\right)_{N\times N},
\end{equation}
where $a_i=\sum_{j\neq i} w_{ij} $. The evolution of the system is determined by the eigenvalues of the Laplacian matrix $\lambda \in \{\lambda_i\} $, which characterize the rate of decay for each mode (eigenvector) of the system as
\begin{equation}\label{EQ:4}
\mathbf{x}(t)=\sum_{i=1}^{N} c_i e^{-\lambda_i t}\mathbf{x}_i,
\end{equation}
where $\mathbf{x}_i$ is the eigenvector of the corresponding Laplacian matrix. Furthermore, the eigenvectors with smaller nonzero eigenvalues decay slowly and control the system's long-term dynamics, and the large eigenvalues vanish rapidly during the early stages of the diffusion process. The spectrum of eigenvalues contains $\lambda_1 = 0$, which is a trivial solution, and the corresponding eigenvector $\mathbf{x_1} = (1,1,\ldots,1)^T$ represents the steady state when $t \rightarrow \infty$. Research on diffusion networks primarily focuses on the second-smallest eigenvalue, which determines the slowest nontrivial mode decay in the system. A larger second eigenvalue indicates faster diffusion, whereas a smaller value implies slower diffusion. Consequently, the weight matrix and its structure affect the eigenvalues and the decay rate of diffusion. 

Studies of diffusion on networks have traditionally focused on finding the second-smallest eigenvalue of the Laplacian matrix to find the slowest nontrivial diffusion mode that determines the convergence rate to equilibrium. In contrast, instead of analyzing individual eigenvalues, this study focused on statistically investigation of eigenvalues to derive conditions under which the network contains fewer fast-decaying diffusion modes. These conditions help us study diffusion dynamics without requiring every eigenvalue and eigenvector. Consequently, this method helps to study the decay of diffusion modes statistically, and provides a computationally efficient framework for estimating diffusion behavior in large-scale networks.  
\section{Model}\label{model}
The model studied in this research describes linear diffusion in a network, as introduced in Eq. ~ (\ref{EQ:1}). The diffusion process in the network continues until the system reaches equilibrium. At the equilibrium state, the quantity stops propagating through the system and becomes the same across all nodes. All elements of the weight matrix are non-negative, $w_{ij} \ge 0$, and remain constant over time. The general solution given in Eq.(\ref{EQ:4}) describes the contributions of the eigenvalues of the Laplacian matrix in Eq.(\ref{EQ:3}). The solution predicts that all nontrivial eigenvalues decay as $t \rightarrow \infty$, and the eigenvector $\mathbf{x_1} = (1,1,\ldots,1)^T$ is the only fixed solution that does not change with time. The following section introduces a statistical framework to investigate the non-trivial eigenvalues for every possible connecting network. This method helps identify the condition under which the network exhibits a reduced number of rapidly decaying diffusion modes.

\section{Analysis}\label{Analysis}
\begin{figure}[t]
	\centering
	\includegraphics[scale = 0.4]{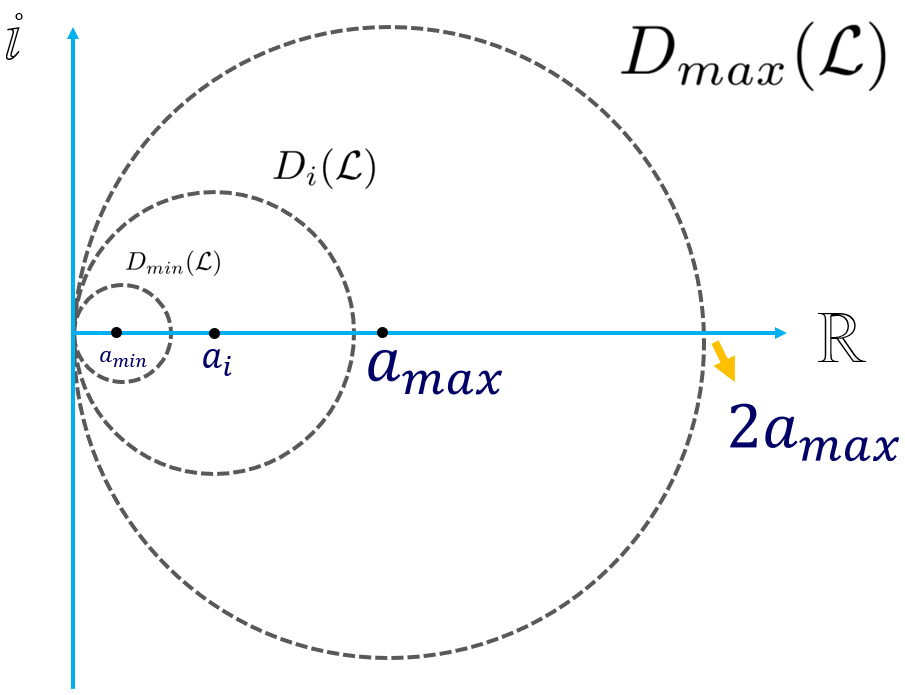}
	\caption{The schematic figure illustrates that the lower bound of the Gershgorin discs is zero, while the upper bound is twice the maximum total weight connected to any node in the network. In this figure, the vertical axis shows the imaginary part of the eigenvalues, i, whereas the horizontal axis represents the real part, R.}
	\label{fig:fig1}
\end{figure}

The first step is to determine the range of possible eigenvalues of the Laplacian matrix introduced in Eq.(\ref{EQ:3}). It needs to use the Gershgorin circle theorem \cite{horn2012matrix} to determine the range of possible eigenvalues for $\mathcal{L} \in \mathbb{C}^{N \times N}$, which represents discs $D_i(\mathcal{L})$ with the radius $R_i=\sum_{j \ne i} \left| \mathcal{L}_{ij} \right|$ and the centroid $C_i=\mathcal{L}_{ii}$ as
\begin{equation}\label{EQ:5}
D_i(\mathcal{L}) \triangleq
\left\{
z \in \mathbb{C}
\,:\,
\left| z - \mathcal{L}_{ii} \right|
\le
\sum_{j \ne i} \left| \mathcal{L}_{ij} \right|
\right\},
\end{equation}
and shows that every possible eigenvalue of the Laplacian matrix lies within at least one of the Gershgorin discs as
\begin{equation}\label{EQ:6}
\lambda_\mathcal{L} \subset \bigcup_{i} D_i(\mathcal{L}).
\end{equation}
For the special structure of the Laplacian matrix, Eq.(\ref{EQ:5}) can be simplified to
\begin{equation}\label{EQ:7}
D_i(\mathcal{L}) \triangleq
\left\{
z \in \mathbb{C}
\,:\,
\left| z - a_i \right|
\le
 a_i
\right\},
\end{equation}
which determines the boundary for all possible eigenvalues as 
\begin{equation}\label{EQ:8}
0 \leq \lambda_\mathcal{L} \leq 2a_{\textbf{max}},
\end{equation}
and Fig.(\ref{fig:fig1}) illustrates a schematic description of the boundary of Gershgorin discs. Moreover, Fig.(\ref{fig:fig1}) shows that the minimum possible eigenvalue is $\lambda_1=0$, which is associated with the trivial eigenvector $\mathbf{x_1} = (1,1,\ldots,1)^T$, and the maximum possible eigenvalue can be twice the maximum value of the total weights connected to a node in the network, which is introduced as $a_{\textbf{max}}=max ( \mathbf{a}=\{a_i\} )$ to simplify the notation. 

In the second step, it can be assumed that the mean of the eigenvalues is known as
\begin{equation}\label{EQ:9}
\mu_\lambda = \frac{1}{N}\sum_{i=1}^N \lambda_i = \frac{1}{N}\operatorname{tr}(\mathcal{L})= \frac{1}{N}\sum_{i=1}^N a_i=\frac{1}{2N}\sum_{i<j}^N w_{ij}  
\end{equation}
, The purpose is to determine the possible range of values for the variance of the eigenspectrum, $\sigma_\lambda^2$. It's trivial that the minimum possible value of the variance of the eigenspectrum is $\sigma_\lambda^2=0$, whereas the maximum value can be obtained from the Bhatia–Davis inequality \cite{bhatia2000better} as
\begin{equation}\label{EQ:10}
\sigma_\lambda^2\leq (2 a_\textbf{max}-\mu_\lambda)(\mu_\lambda).
\end{equation}
The final step is to determine the maximum probability of finding a Laplacian matrix eigenvalue greater than $k$ for $k \ge \mu$ or the maximum probability of finding a Laplacian matrix eigenvalue less than $k$ for $k \leq\mu$. In this section, the upper bound is derived using Cantelli's inequality \cite{cantelli1929} as
 \begin{equation}\label{EQ:11}
\operatorname{Pr}(\lambda\ge k) \leq\frac{\sigma_\lambda^2}{\sigma_\lambda^2+(k-\mu_\lambda)^2}
\end{equation}
, and by using the Eq.(\ref{EQ:10}), the Eq.(\ref{EQ:11}) can be written as 
 \begin{equation}\label{EQ:12}
\operatorname{Pr}(\lambda\ge k) \leq\frac{\mu_\lambda(2 a_\textbf{max}-\mu_\lambda)}{\mu_\lambda(2 a_\textbf{max}-\mu_\lambda)+(k-\mu_\lambda)^2}
\end{equation}
, and the lower bound of the maximum probability of finding a Laplacian matrix eigenvalue less than $k$ for $ k \leq \mu$ is 
 \begin{equation}\label{EQ:13}
\operatorname{Pr}(\lambda \leq k) \leq\frac{\mu_\lambda(2 a_\textbf{max}-\mu_\lambda)}{\mu_\lambda(2 a_\textbf{max}-\mu_\lambda)+(\mu_\lambda - k)^2}.
\end{equation}

\begin{figure}[t]
	\centering
	\includegraphics[scale = 0.350]{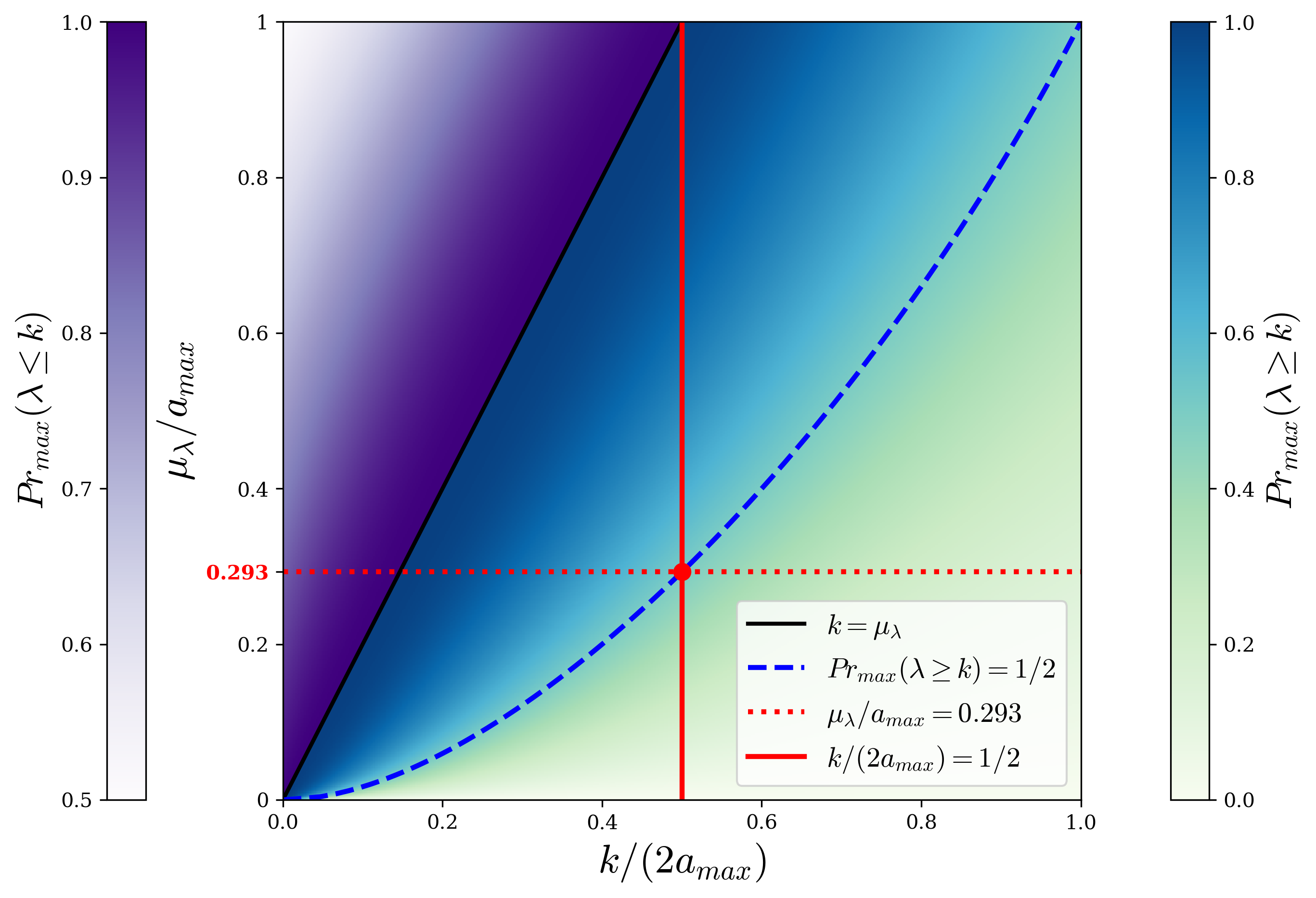}
	\caption{The figure shows the maximum probability of finding eigenvalues greater than $k$ for $k \ge \mu$, and less than $k$ for $k \leq \mu$. The blue dashed line represents the maximum probability equal to $\frac{1}{2}$; the intersection with the red line gives the condition ensuring that, in the upper half of the eigenvalue subspace, the probability of finding an eigenvalue is no greater than one-half.} 
	\label{fig:fig2}
\end{figure}

The idea is to determine the value of $\mu_\lambda$ that ensures the probability of finding eigenvalues greater than $a_\textbf{max}$ is equal to or less than $\frac{1}{2}$. This condition states that eigenvalues are more likely to exist in the lower half of the possible eigenvalues subspace. The considered $\mu_\lambda$ is the answer to the following equation 
 \begin{equation}\label{EQ:14}
\frac{\mu_\lambda(2 a_\textbf{max}-\mu_\lambda)}{\mu_\lambda(2 a_\textbf{max}-\mu_\lambda)+(a_\textbf{max} -\mu_\lambda)^2} = \frac{1}{2},
\end{equation}
the solution of Eq.(\ref{EQ:14}) is $\mu_\lambda=a_\textbf{max}(1\pm \frac{\sqrt{2}}{2})$, indicates that the $0 \leq \mu_\lambda \leq a_\textbf{max}$ is acceptable as the Eq.(\ref{EQ:8}) implies all of the eigenspectrum is positive. Consequently, $\mu_\lambda=a_\textbf{max}(1 - \frac{\sqrt{2}}{2})=0.2929a_\textbf{max}$ is the only acceptable answer, as it ensures that in the upper half of the eigenvalue subspace, the probability of finding an eigenvalue is no greater than one-half, which is represented in Fig.(\ref{fig:fig2}). Furthermore, it's valuable to know how the maximum probability of finding eigenvalues greater than or less than $k$ changes with a small difference in network connections. It's convenient to define the maximum probability associated with Eq.(\ref{EQ:12}) as
 \begin{equation}\label{EQ:15}
\operatorname{Pr_{max}}(\lambda\ge k) =\frac{\mu_\lambda(2 a_\textbf{max}-\mu_\lambda)}{\mu_\lambda(2 a_\textbf{max}-\mu_\lambda)+(k-\mu_\lambda)^2}
\end{equation}
and, similarly, the maximum probability associated with Eq.(\ref{EQ:13}) is
 \begin{equation}\label{EQ:16}
\operatorname{Pr_{max}}(\lambda \leq k) =\frac{\mu_\lambda(2 a_\textbf{max}-\mu_\lambda)}{\mu_\lambda(2 a_\textbf{max}-\mu_\lambda)+(\mu_\lambda - k)^2}.
\end{equation}
To simplify the calculation. The first part of the calculation involves determining the derivative of $\operatorname{Pr_{max}}= \operatorname{Pr_{max}}(\lambda\ge k) = \operatorname{Pr_{max}}(\lambda\leq k )$ with respect to $w_{ij}$, when $w_{ij}$ does not belong to the row corresponding to $a_\textbf{max}$, which is calculated as

\begin{figure}[t]
	\centering
	\includegraphics[scale = 0.350]{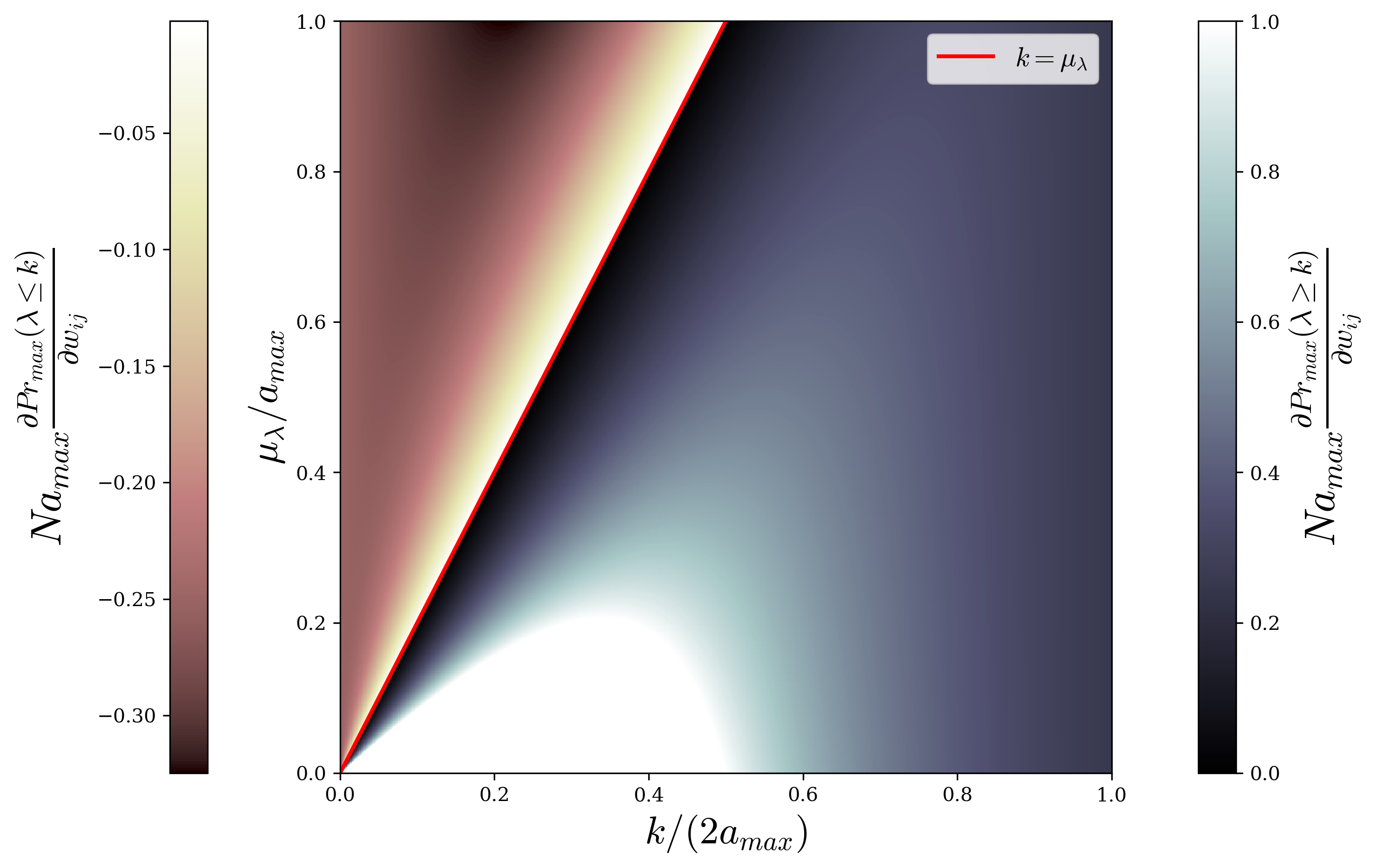}
	\caption{The figure represents the derivative of the maximum probability with respect to $w_{ij}$ when $w_{ij}$ does not belong to the row corresponding to $a_\textbf{max}$. The derivative is multiplied by $Na_\textbf{max}$ to remove the dependence on the factor $\frac{1}{Na_\textbf{max}}$. For $k \geq \mu$, the derivative is plotted over the range from 0 to 1 because it diverges at the origin $(0,0)$.} 
	\label{fig:fig3}
\end{figure}

\begin{figure}[t]
	\centering
	\includegraphics[scale = 0.370]{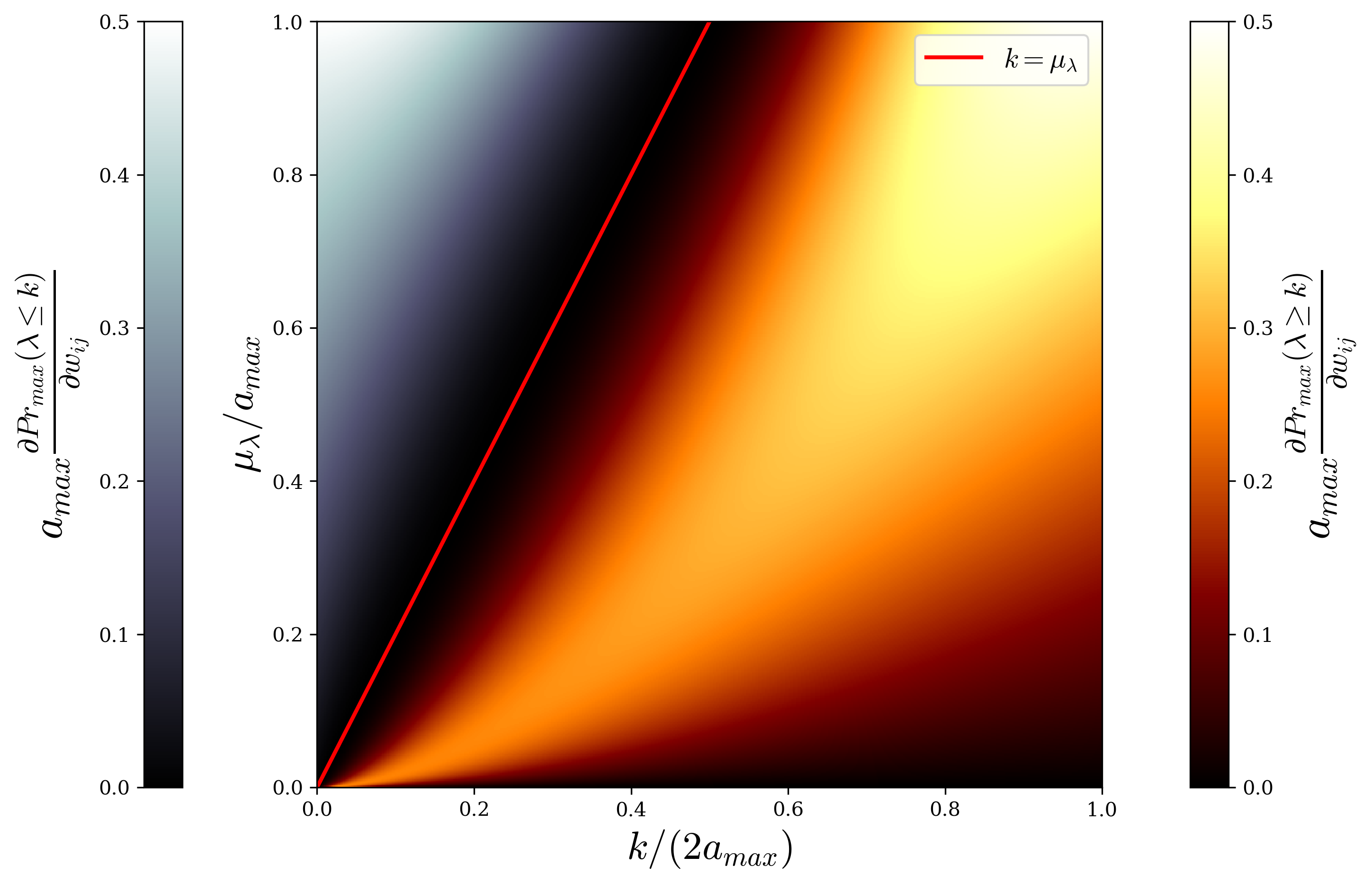}
	\caption{The figure represents the derivative of the maximum probability with respect to $w_{ij}$ when $w_{ij}$ belongs to the row corresponding to $a_\textbf{max}$. The derivative is multiplied by $a_\textbf{max}$ to eliminate its dependence on the factor $\frac{1}{a_\textbf{max}}$.} 
	\label{fig:fig4}
\end{figure}

\begin{figure}[t]
	\centering
	\includegraphics[scale = 0.370]{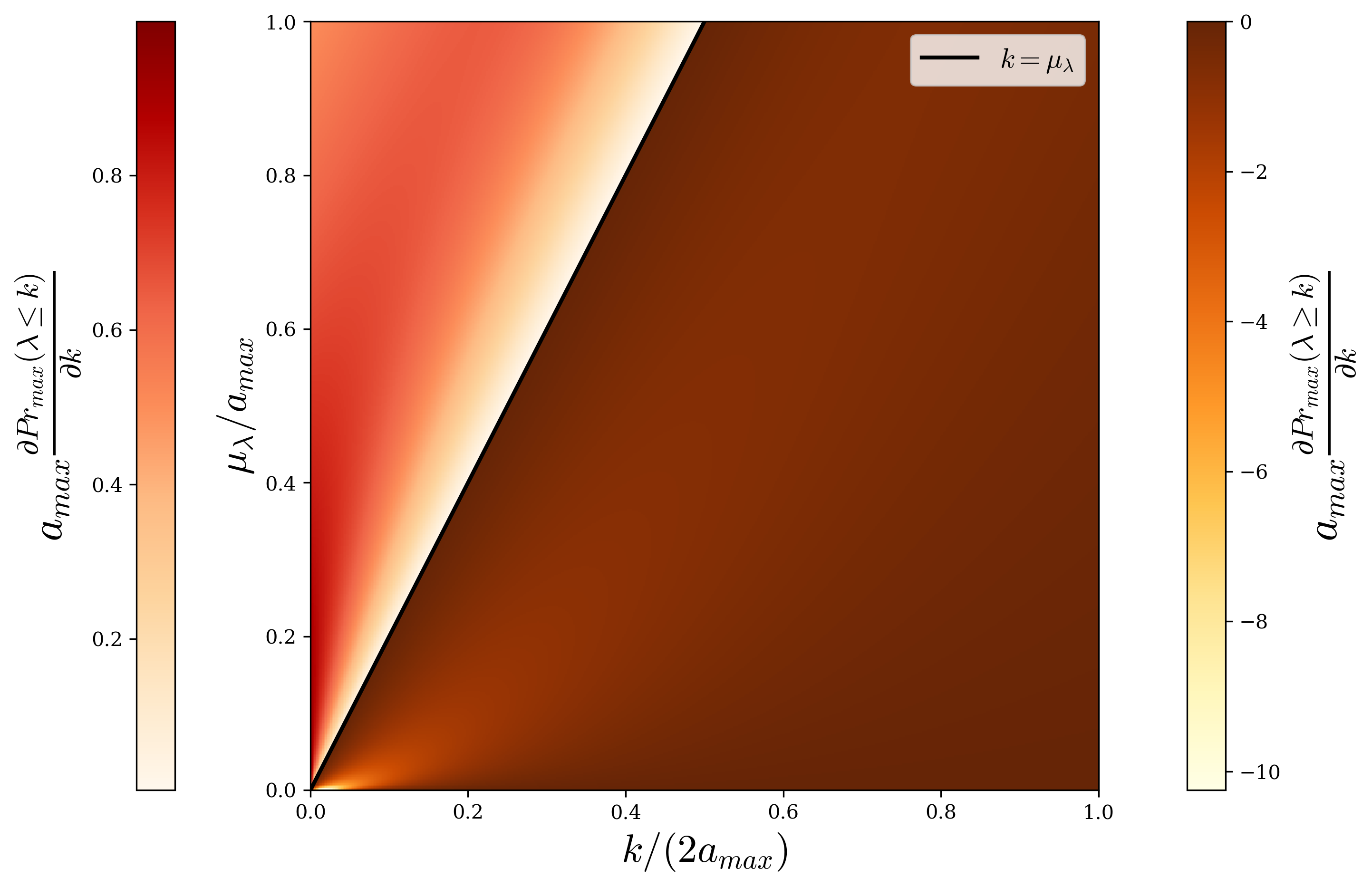}
	\caption{The figure shows the derivative of the maximum probability with respect to $ k $. The derivative is multiplied by $a_\textbf{max}$ to remove its dependence on the factor $\frac{1}{a_\textbf{max}}$.} 
	\label{fig:fig5}
\end{figure}

\begin{widetext}
\begin{equation}\label{EQ:17}
\frac{\partial}{\partial w_{ij}}
\left(
\operatorname{Pr_{max}}
\right)
=
\frac{
\left(2(a_{\max}-k)\mu_\lambda+k^2\right)
\frac{(a_{\max}-\mu_\lambda)}{N}
-
\left(2a_{\max}\mu_\lambda-\mu_\lambda^2\right)
\frac{(a_{\max}-k)}{N}
}
{\left(2(a_{\max}-k)\mu_\lambda+k^2\right)^2},
\end{equation} 
 Eq.(\ref{EQ:17}) shows that $\frac{\partial}{\partial w_{ij}} \left(Pr_{max}\right)$ is proportional to $\frac{1}{N}$. Therefore, for a finite value of $a_\textbf{max}$, this derivative vanishes in the limit $N \rightarrow \infty$. To obtain a non-vanishing quantity in this limit, it is more appropriate to consider $N\frac{\partial}{\partial w_{ij}} \left(Pr_{max}\right)$ as
\begin{equation}\label{EQ:18}
N\frac{\partial}{\partial w_{ij}}
\left(
\operatorname{Pr_{max}}
\right)
=
\frac{
\left(2(a_{\max}-k)\mu_\lambda+k^2\right)
(a_{\max}-\mu_\lambda)
-
\left(2a_{\max}\mu_\lambda-\mu_\lambda^2\right)(a_{\max}-k)}
{\left(2(a_{\max}-k)\mu_\lambda+k^2\right)^2}.
\end{equation} 
After applying transformation $\mu \rightarrow\frac{\mu}{a_{max}}$ and $k \rightarrow\frac{k}{2a_{max}}$, the derivative of the probability scales as $\frac{1}{a_{max}}$. Fig.(\ref{fig:fig3}) shows $N a_{max}\frac{\partial\operatorname{Pr_{max}}(\lambda \ge k)}{\partial w_{ij}}$ and $N a_{max} \frac{\partial\operatorname{Pr_{max}}(\lambda \leq k)}{\partial w_{ij}}$, removing the overall factor $\frac{1}{N a_{max}}$.

Next, it is considered the case where $w_{ij}$ belongs to the row corresponding to $a_{\max}$, and the derivative of $\operatorname{Pr_{max}}$ is then given by
\begin{equation}\label{EQ:19}
\frac{\partial}{\partial w_{ij}}
\left(
Pr_{max}
\right)
=
\frac{
\left(2(a_{\max}-k)\mu_\lambda+k^2\right)
\left(
2\mu_\lambda+\frac{a_{\max}-\mu_\lambda}{N}
\right)
-
\left(2a_{\max}\mu_\lambda-\mu_\lambda^2\right)
\left(
2\mu_\lambda+\frac{a_{\max}-k}{N}
\right)
}
{\left(2(a_{\max}-k)\mu_\lambda+k^2\right)^2},
\end{equation}
which in the limit $N \rightarrow \infty$, and for finite $a_\textbf{max}$, the Eq.(\ref{EQ:19}) convert to
\begin{equation}\label{EQ:20}
\frac{\partial}{\partial w_{ij}}
\left(
Pr_{max}
\right)
=
\frac{
2\mu_\lambda\left(2(a_{\max}-k)\mu_\lambda+k^2\right)
-
2\mu_\lambda\left(2a_{\max}\mu_\lambda-\mu_\lambda^2\right)
}
{\left(2(a_{\max}-k)\mu_\lambda+k^2\right)^2},
\end{equation}
After applying transformation of $\mu \rightarrow\frac{\mu}{a_{max}}$ and $k \rightarrow\frac{k}{2a_{max}}$, derivative of the probability scales as $\frac{1}{a_{max}}$. Fig.(\ref{fig:fig4}) represents $a_{max} \frac{\partial\operatorname{Pr_{max}}(\lambda \ge k)}{\partial w_{ij}}$ and $a_{max} \frac{\partial\operatorname{Pr_{max}}(\lambda \leq k)}{\partial w_{ij}}$, which removes the dependence on the factor $\frac{1}{ a_{max}}$.

It's also useful to examine how the maximum probability of finding eigenvalues greater than or less than $k$, as defined in Eqs. (\ref{EQ:15}) and (\ref{EQ:16}), changes as $k$ is slightly varied. To investigate the sensitivity of $\operatorname{Pr_{max}}$ to changes in $k$, its derivative with respect to $k$ is calculated as
\begin{equation}\label{EQ:21}
\frac{\partial \Pr_{\max}}{\partial k}
=
\frac{
2\mu_\lambda(2a_{\max}-\mu_\lambda)(\mu_\lambda- k)
}{
\left[
\mu_\lambda(2a_{\max}-\mu_\lambda)+(k-\mu_\lambda)^2
\right]^2
},
\end{equation}
the transformations $\mu \rightarrow\frac{\mu}{a_{max}}$ and $k \rightarrow\frac{k}{2a_{max}}$ makes the derivative of probability proportional to $\frac{1}{a_{max}}$. Therefore, Fig.(\ref{fig:fig5}) illustrates $a_{max} \frac{\partial\operatorname{Pr_{max}}(\lambda \ge k)}{\partial w_{ij}}$ and $a_{max} \frac{\partial\operatorname{Pr_{max}}(\lambda \leq k)}{\partial w_{ij}}$ to eliminate the dependence on the factor $\frac{1}{ a_{max}}$.
\end{widetext}

\section{Conclusion}
In conclusion, researchers have usually studied diffusion on networks using the exact eigenvalues of the Laplacian matrix. Most earlier studies have focused on the second-smallest eigenvalue to find the slowest decay rate and the timescale for the system to reach equilibrium. However, computing the exact eigenvalues is difficult because the number of nodes in the network increases. In this study, instead of calculating the exact eigenvalues, I studied the Laplacian eigenspectrum statistically in the limit $N \rightarrow \infty$. Studying the statistical properties of the Laplacian matrix is helpful to better understand the behaviour of the network in diffusion process , especially when the system is not close to the steady state. I used the Gershgorin circle theorem to determine the possible range of eigenvalues, while the Bhatia–Davis inequality provides an upper bound for the variance of the eigenvalue distribution. Moreover, I used Cantelli’s inequality to determine the maximum probability of finding eigenvalues greater than k when $\mu \leq k$ or less than k when $k \leq \mu$. Using these results, I found the condition $\mu = 0.2929 a_{max}$ that ensures the system is less diffusive. Here, the system is called "less diffusive", meaning that in the space of possible eigenvalues, more than half of the eigenvalues lie in the lower bound of the space. Consequently, A larger fraction of the modes decay more slowly because the eigenvalues influence the decay rates of the diffusion modes. The condition $\mu = 0.2929a_{\max}$ shows that the distribution of the eigenvalues depends on the ratio between the average value of $a_i$ and its maximum value, where $a_i$ represents the total weight connected to node $i$.

In the second section, I investigate the derivatives of $\operatorname{Pr_{max}}$ with respect to $w_{ij}$ to show how changing a network weight influences the eigenvalue spectrum, and how the network responds when $k$ changes. Fig.(\ref{fig:fig3}) represents the derivative of $\operatorname{Pr_{max}}= \operatorname{Pr_{max}}(\lambda\ge k) = \operatorname{Pr_{max}}(\lambda\leq k )$ with respect to $w_{ij}$, when $w_{ij}$ does no belong to the row corresponding to $a_\textbf{max}$. the derivative of $\operatorname{Pr_{max}}(\lambda\leq k )$ declines when $w_{ij}$ increases. Moreover, the result indicates that when $w_{ij}$ does not belong to $a_{max}$, a small change in $w_{ij}$ increases $\operatorname{Pr_{max}}(\lambda \ge k)$, and it goes to infinity when $k \rightarrow 0$; therefore, the figure Fig.~(\ref{fig:fig3}) is shown over the range zero to one to keep it readable. When $w_{ij}$ belongs to the row corresponding to $a_{\mathbf{max}}$, the behavior is different because changing $w_{ij}$ affects both the mean eigenvalue $\mu$ and the upper bound of the eigenspectrum, $a_{max}$. One term of the resulting derivatives $\operatorname{Pr_{max}}(\lambda \ge k)$ and $\operatorname{Pr_{max}}(\lambda \leq k)$ based on $w_{ij}$ is proportional to $\frac{1}{N}$, and when $N \rightarrow \infty$ this part vanish. After taking this limit, the remaining terms show that the derivatives of both $\operatorname{Pr_{max}}(\lambda \ge k)$ and $\operatorname{Pr_{max}}(\lambda \leq k)$ based on $w_{ij}$ are positive and finite. The result shows that even in the regime $k \leq \mu$, increasing $w_{ij}$ can increase $\operatorname{Pr}_{max}(\lambda\leq k)$. This result is not trivial and shows that when $w_{ij}$ contributes to $a_{\max}$, increasing the upper bound of the eigenspectrum has a stronger effect on the probability than the increase in $\mu$. Therefore, the location of a weight in the network, particularly whether it contributes to the node with the largest total weight, plays an important role in determining how changes in individual weights affect the eigenspectrum. I calculated the derivative of $\operatorname{Pr_{max}}$ with respect to $k$ to show how the maximum probability is sensitive when the threshold changes. Since $k$ determines the boundary separating the eigenspectrum into the regions above and below the selected threshold, this derivative indicates how strongly the probability changes when the boundary is slightly shifted.

Finally, several directions remain for extending the method. For example, Cantelli's and Bhatia–Davis inequalities can be replaced with sharper bounds to obtain tighter estimates of the maximum probability of finding eigenvalues greater than $k$, improving the accuracy of the results. Furthermore, the statistical analysis of the eigenspectrum of the Laplacian matrix is not limited to the method introduced in this study, and there are several methods to use statistics to study the modes in a diffusion network without requiring the exact calculation of all eigenvalues. This method provides more computationally efficient ways to analyze diffusion dynamics in large-scale networks.
\bibliography{MyReferences}

@article{barabasi2013network,
  title={Network science},
  author={Barab{\'a}si, Albert-L{\'a}szl{\'o}},
  journal={Philosophical Transactions of the Royal Society A: Mathematical, Physical and Engineering Sciences},
  volume={371},
  number={1987},
  pages={20120375},
  year={2013},
  publisher={The Royal Society Publishing}
}

@article{barabasi1999emergence,
  title={Emergence of scaling in random networks},
  author={Barab{\'a}si, Albert-L{\'a}szl{\'o} and Albert, R{\'e}ka},
  journal={science},
  volume={286},
  number={5439},
  pages={509--512},
  year={1999},
  publisher={American Association for the Advancement of Science}
}

@article{albert2002statistical,
  title={Statistical mechanics of complex networks},
  author={Albert, R{\'e}ka and Barab{\'a}si, Albert-L{\'a}szl{\'o}},
  journal={Reviews of modern physics},
  volume={74},
  number={1},
  pages={47},
  year={2002},
  publisher={APS}
}

@article{boccaletti2006complex,
  title={Complex networks: Structure and dynamics},
  author={Boccaletti, Stefano and Latora, Vito and Moreno, Yamir and Chavez, Martin and Hwang, D-U},
  journal={Physics reports},
  volume={424},
  number={4-5},
  pages={175--308},
  year={2006},
  publisher={Elsevier}
}

@book{newman2018networks,
  title={Networks},
  author={Newman, Mark},
  year={2018},
  publisher={Oxford university press}
}

@article{milgram1967small,
  title={The small world problem},
  author={Milgram, Stanley and others},
  journal={Psychology today},
  volume={2},
  number={1},
  pages={60--67},
  year={1967},
  publisher={New York}
}

@article{watts1998collective,
  title={Collective dynamics of ‘small-world’networks},
  author={Watts, Duncan J and Strogatz, Steven H},
  journal={nature},
  volume={393},
  number={6684},
  pages={440--442},
  year={1998},
  publisher={Nature Publishing Group}
}

@article{jeong1999diameter,
  title={Diameter of the world-wide web},
  author={Jeong, AH},
  journal={Nature},
  volume={401},
  pages={130--131},
  year={1999}
}

@article{faloutsos1999power,
  title={On power-law relationships of the internet topology},
  author={Faloutsos, Michalis and Faloutsos, Petros and Faloutsos, Christos},
  journal={ACM SIGCOMM computer communication review},
  volume={29},
  number={4},
  pages={251--262},
  year={1999},
  publisher={ACM New York, NY, USA}
}

@article{jeong2000large,
  title={The large-scale organization of metabolic networks},
  author={Jeong, Hawoong and Tombor, B{\'a}lint and Albert, R{\'e}ka and Oltvai, Zoltan N and Barab{\'a}si, A-L},
  journal={Nature},
  volume={407},
  number={6804},
  pages={651--654},
  year={2000},
  publisher={Nature Publishing Group UK London}
}

@article{jeong2001lethality,
  title={Lethality and centrality in protein networks},
  author={Jeong, Hawoong and Mason, Sean P and Barab{\'a}si, A-L and Oltvai, Zoltan N},
  journal={Nature},
  volume={411},
  number={6833},
  pages={41--42},
  year={2001},
  publisher={Nature Publishing Group UK London}
}

@article{mohar1991laplacian,
  title={The Laplacian spectrum of graphs},
  author={Mohar, Bojan and Alavi, Y and Chartrand, G and Oellermann, Ortrud},
  journal={Graph theory, combinatorics, and applications},
  volume={2},
  number={871-898},
  pages={12},
  year={1991},
  publisher={Wiley}
}

@article{raj2012network,
  title={A network diffusion model of disease progression in dementia},
  author={Raj, Ashish and Kuceyeski, Amy and Weiner, Michael},
  journal={Neuron},
  volume={73},
  number={6},
  pages={1204--1215},
  year={2012},
  publisher={Elsevier}
}

@article{abdelnour2014network,
  title={Network diffusion accurately models the relationship between structural and functional brain connectivity networks},
  author={Abdelnour, Farras and Voss, Henning U and Raj, Ashish},
  journal={Neuroimage},
  volume={90},
  pages={335--347},
  year={2014},
  publisher={Elsevier}
}

@article{pastor2015epidemic,
  title={Epidemic processes in complex networks},
  author={Pastor-Satorras, Romualdo and Castellano, Claudio and Van Mieghem, Piet and Vespignani, Alessandro},
  journal={Reviews of modern physics},
  volume={87},
  number={3},
  pages={925--979},
  year={2015},
  publisher={APS}
}

@article{keeling2005networks,
  title={Networks and epidemic models},
  author={Keeling, Matt J and Eames, Ken TD},
  journal={Journal of the royal society interface},
  volume={2},
  number={4},
  pages={295},
  year={2005}
}

@article{gomez2012inferring,
  title={Inferring networks of diffusion and influence},
  author={Gomez-Rodriguez, Manuel and Leskovec, Jure and Krause, Andreas},
  journal={ACM Transactions on Knowledge Discovery from Data (TKDD)},
  volume={5},
  number={4},
  pages={1--37},
  year={2012},
  publisher={ACM New York, NY, USA}
}

@article{bouchet2026directionality,
  title={Directionality-induced jamming in multiplex networks},
  author={Bouchet, Mateo and Tejedor, Alejandro and Wang, Xiangrong and Moreno, Yamir},
  journal={Physical Review Letters},
  volume={136},
  number={20},
  pages={207401},
  year={2026},
  publisher={APS}
}

@article{gomez2013diffusion,
  title={Diffusion dynamics on multiplex networks},
  author={Gomez, Sergio and Diaz-Guilera, Albert and Gomez-Gardenes, Jesus and Perez-Vicente, Conrad J and Moreno, Yamir and Arenas, Alex},
  journal={Physical review letters},
  volume={110},
  number={2},
  pages={028701},
  year={2013},
  publisher={APS}
}

@article{gomez2008entropy,
  title={Entropy rate of diffusion processes on complex networks},
  author={G{\'o}mez-Gardenes, Jes{\'u}s and Latora, Vito},
  journal={Physical Review E—Statistical, Nonlinear, and Soft Matter Physics},
  volume={78},
  number={6},
  pages={065102},
  year={2008},
  publisher={APS}
}

@article{pastor2001epidemic,
  title={Epidemic spreading in scale-free networks},
  author={Pastor-Satorras, Romualdo and Vespignani, Alessandro},
  journal={Physical review letters},
  volume={86},
  number={14},
  pages={3200},
  year={2001},
  publisher={APS}
}

@article{serrano2006,
  title={Clustering in complex networks. II. Percolation properties},
  author={Serrano, M {\'A}ngeles and Bogun{\'a}, Mari{\'a}n},
  journal={Physical Review E—Statistical, Nonlinear, and Soft Matter Physics},
  volume={74},
  number={5},
  pages={056115},
  year={2006},
  publisher={APS}
}

@article{nematzadeh2014optimal,
  title={Optimal network modularity for information diffusion},
  author={Nematzadeh, Azadeh and Ferrara, Emilio and Flammini, Alessandro and Ahn, Yong-Yeol},
  journal={arXiv preprint arXiv:1401.1257},
  year={2014}
}

@article{watts2002simple,
  title={A simple model of global cascades on random networks},
  author={Watts, Duncan J},
  journal={Proceedings of the National Academy of Sciences},
  volume={99},
  number={9},
  pages={5766--5771},
  year={2002},
  publisher={The National Academy of Sciences}
}

@article{nakao2010turing,
  title={Turing patterns in network-organized activator--inhibitor systems},
  author={Nakao, Hiroya and Mikhailov, Alexander S},
  journal={Nature Physics},
  volume={6},
  number={7},
  pages={544--550},
  year={2010},
  publisher={Nature Publishing Group UK London}
}

@article{castellano2009nonlinear,
  title={Nonlinear q-voter model},
  author={Castellano, Claudio and Mu{\~n}oz, Miguel A and Pastor-Satorras, Romualdo},
  journal={Physical Review E—Statistical, Nonlinear, and Soft Matter Physics},
  volume={80},
  number={4},
  pages={041129},
  year={2009},
  publisher={APS}
}

@book{horn2012matrix,
  title={Matrix analysis},
  author={Horn, Roger A and Johnson, Charles R},
  year={2012},
  publisher={Cambridge university press}
}

@article{bhatia2000better,
  title={A better bound on the variance},
  author={Bhatia, Rajendra and Davis, Chandler},
  journal={The american mathematical monthly},
  volume={107},
  number={4},
  pages={353--357},
  year={2000},
  publisher={Taylor \& Francis}
}

@inproceedings{cantelli1929,
  author    = {Cantelli, Francesco Paolo},
  title     = {Sui confini della probabilit{\`a}},
  booktitle = {Atti del Congresso Internazionale dei Matematici},
  pages     = {47--60},
  year      = {1929},
  address   = {Bologna}
}
\end{document}